\documentclass[fp,twocolumn]{jpsj3}
\usepackage{txfonts}
\usepackage[usenames, dvipsnames]{color}
\usepackage[flushleft]{threeparttable}
\usepackage{epstopdf}
\newcommand{\beq}{\begin{eqnarray}}
\newcommand{\eeq}{\end{eqnarray}}

\title{First-principles Study of Spin-wave Excitations of 3$d$ Transition Metals with Linear Combination of Pseudo-atomic Orbitals}

\author{Teguh Budi Prayitno$^{1}$\thanks{teguh-budi@unj.ac.id} and Fumiyuki Ishii$^2$}
\inst{$^1$Department of Physics, Faculty of Mathematics and Natural Science, Universitas Negeri Jakarta, Kampus A Jl. Rawamangun Muka, Jakarta Timur 13220, Indonesia\\
$^2$Nanomaterials Research Institute, Kanazawa University, Kanazawa 920-1192, Japan} 

\abst{We have employed the generalized Bloch theorem to evaluate the spin stiffness constants of 3$d$ transition metals (bcc-Fe, fcc-Co, and fcc-Ni) within the linear combination of pseudo-atomic orbitals (LCPAO). The spin stiffness constants were obtained by fitting the spin-wave energy curve, which relates to the total energy difference and the spiral vectors. In order to convince the reliable spin stiffness constants, we also provided the convergences of spin stiffness constants in terms of the cutoff radius and the number of orbitals. After observing the specific cutoff radius and the basis orbital, at which the spin stiffness constant converges, we used those two parameters to compute the Curie temperature by using the mean field approximation and the random phase approximation. For the latter approximation, we applied the so-called Debye approximation, which is intended to reduce very significantly many required wavevectors to evaluate the Curie temperature. We claimed that our results are in good agreement with both other calculations and experiments.  
}

\begin{document}
\maketitle

\section{Introduction}
The recent study based on the magnetic structures within the itinerant electron model gives a strong description of why some metals should have a ferromagnetic ground state. This ground state can be well-explained by the so-called Stoner criterion, which is determined by the Coulomb interaction. The Stoner criterion has successfully predicted that some 3$d$ transition metals, such as bcc-Fe, fcc-Co, and fcc-Ni, have the ferromagnetic ground state. The next task is to investigate the magnetic excitations and to evaluate the Curie temperature of these itinerant ferromagnets. 

The excitations in magnetic systems are usually addressed to two different kinds of excitations. The first excitations are called the Stoner excitations. These excitations do exist due to the transfer of electron-hole, namely, the transition of an electron from the filled state to the empty state. Due to the excitations of electron-hole, the continuum region (Stoner continuum) is created along the certain wavevector. The second ones are the spin-wave excitations (magnons), where the configuration of magnetic moments of all atoms gives a spiral form. The clear difference lies in the range of the wavelength spectra. Unlike the Stoner excitations, the spin-wave excitations only hold for the short wavevector (long wavelength) and are then damped when entering the Stoner continuum. Nevertheless, the spin-wave excitations become much more dominant in the nearly lowest excitations spectrum. This means that the Stoner excitations can be excluded up to the critical temperature on this condition. This approach enables us to estimate the Curie temperature by considering the spin-wave excitations.           

There are two approaches to consider the spin-wave excitations, i.e., the real space approach and the reciprocal space approach. In the real space approach, one should first calculate the exchange coupling constant of two different atoms to obtain the spin-wave energy. On the contrary, the spin-wave energy can be directly calculated by using the reciprocal space method, which implements the so-called generalized Bloch theorem (GBT). The limitations of these approaches lie in the efficiency of some calculations. According to Padja et al. \cite{1Padja}, the more efficient calculations of the exchange coupling constant or the Curie temperature can be performed within the real space approach than the reciprocal space approach. Contrarily, the spin-wave energy or the spin stiffness constant can be computed more efficiently using the reciprocal space approach.

So far, we only note that the use of GBT with an LCPAO was never used by the previous authors to study the spin-wave excitations, especially for calculating the Curie temperature, in the framework of first-principles approach for the ferromagnetic 3$d$ transition metals \cite{1Padja,2Pindor, 3Liech, Sabiryanov, 4Staunt, 5Uhl, 6Halilov1, 7Rosengaard, 8Halilov2, 9Shall, 10Kubler}. Most calculations of the spin-wave excitations are performed by using the plane-wave basis sets, which give good accuracy. Nevertheless, the computational cost is very high due to a large number of plane waves, especially for a system having a vacuum region. At the same situation, an LCPAO can give the same accuracy without a large number of basis sets. Since all methods have their own treatments to study the spin-wave excitations, we propose our LCPAO to calculate the spin stiffness constants of 3$d$ transition metals (bcc-Fe, fcc-Co, and fcc-Ni) based on the converged results with respect to the cutoff radius and the number of orbitals as a basis set. This treatment was successfully applied to predict some physical quantities for some systems \cite{11Ozaki, 12Ozaki, 13Yoon}. We also show that the appropriate number of orbitals depends on the cutoff radii to obtain reliable results. These cutoff radii are the parameter as the boundary conditions to create the pseudo-atomic orbitals (PAO) through the confinement scheme \cite{11Ozaki, 12Ozaki}. 

In this study, we apply the GBT with the constraint scheme method to calculate the magnon energy using the reciprocal space method (frozen magnon method) within the density functional theory (DFT). Our electronic calculation, in which the wavefunction of a single particle is described by an LCPAO with the norm-conserving pseudopotentials, reproduces successfully the spin stiffness constants and Curie temperatures of 3$d$ transition metals (bcc-Fe, fcc-Co, and fcc-Ni). We also show that the magnon dispersion relations for these systems on the high-symmetry lines in the Brillouin zone are good agreement with the previous references. Regarding the Curie temperature, we apply the so-called Debye approximation to reduce many required wavevectors within the random phase approximation (RPA). In a different way, we also estimate the Curie temperature within the mean field approximation (MFA) by calculating the average value of the magnon energies in the Brillouin zone. We believe that our discussions using an LCPAO gives more completed results than those using the other methods.           

\section{Mathematical Formulation}

We perform the first-principles approach on the magnetic excitations of 3$d$ transition metals using the OpenMX code \cite{14Openmx}, which uses the norm-conserving pseudopotentials \cite{15Troullier} and an LCPAO as the basis set \cite{11Ozaki, 12Ozaki}. In the Openmx code, the wavefunction of a single particle in the noncollinear DFT, which implements the GBT, is described by an LCPAO on the site $\tau_{i}$ as \cite{16Teguh} 
 \beq
\psi_{\nu\textit{\textbf{k}}}\left(\textit{\textbf{r}}\right)&=&\frac{1}{\sqrt{N}}\sum_{n}^{N}\sum_{i\alpha}\left[e^{i\left(\textit{\textbf{k}}-\frac{\textit{\textbf{q}}}{2}\right)\cdot\textit{\textbf{R}}_{n}}C_{\nu\textit{\textbf{k}},i\alpha}^{\uparrow}
\left(
\begin{array}{cc}
1\\
0\end{array}
\right)\right.\nonumber\\
& &\left.+e^{i\left(\textit{\textbf{k}}+\frac{\textit{\textbf{q}}}{2}\right)\cdot\textit{\textbf{R}}_{n}}C_{\nu\textit{\textbf{k}},i\alpha}^{\downarrow}\left(
\begin{array}{cc}
0\\
1\end{array}
\right)\right]\nonumber\\
& &\times\phi_{i\alpha}\left(\mathrm{\textit{\textbf{r}}-\tau_{i}-\textit{\textbf{R}}_{n}}\right).\label{lcpao}
\eeq  
where $\phi_{i\alpha}$ denotes the PAO and $\textit{\textbf{q}}$ is the reciprocal spiral vector. The similar formulation can also be found in Ref. \citen{17Garcia, 18Garcia}, which also implements the GBT in the code using an LCPAO \cite{19Siesta}. In these references, the authors only implemented the GBT to investigate the spiral ground state of $\gamma$-Fe for several lattice constants. It means that the implementation of the GBT to discuss the spin-wave excitations has not been explored yet. 

The implication of the wavefunction in Eq. (\ref{lcpao}) is that the magnetic moment of each magnetic atom will be rotated along the chosen spiral vector $\textit{\textbf{q}}$ from one cell to the other cells, which is given by \cite{20Sandratskii}
   \beq
	\textit{\textbf{M}}_{i}(\textit{\textbf{r}}+\textit{\textbf{R}}_{i})=M_{i}(\textit{\textbf{r}}) \left(
\begin{array}{cc}
\cos\left(\varphi_{0}+\textit{\textbf{q}}\cdot \textit{\textbf{R}}_{i}\right)\sin\theta_{i}\\
\sin\left(\varphi_{0}+\textit{\textbf{q}}\cdot \textit{\textbf{R}}_{i}\right)\sin\theta_{i}\\
\cos\theta_{i}\end{array} 
\right). \label{moment}  
\eeq
Here, $\theta$ is defined as the cone angle, which should be fixed to discuss the spin-wave excitations. Meanwhile, the azimuthal angle $\varphi$ is rotated by the scalar product between $\textit{\textbf{q}}$ and the lattice vector $\textit{\textbf{R}}$. The constraint scheme method is further applied to fix the direction of the magnetic moment, for example see Refs. \citen{21Gebauer, 22Kurz}. Through the above formulas, we carry out the DFT calculation of the total energy for each $\textit{\textbf{q}}$ by using the primitive unit cell and neglecting the spin orbit interaction (SOI).

Since the Heisenberg model suits to investigate the long wavelength spin-wave excitations, the calculated total energy in DFT calculation should be mapped onto the energy in the Heisenberg Hamiltonian. If the two magnetic moments interact with each other characterized by the exchange coupling constant $J_{ij}$, the total energy of the system can be written in terms of the Heisenberg Hamiltonian 
		\beq
E&=&E\left(M_{i}^{2}\right) -\frac{1}{2N} \sum_{i\neq j}J_{ij}\textit{\textbf{M}}_{i}\cdot\textit{\textbf{M}}_{j}\nonumber\\
     &=& E\left(M_{i}^{2}\right)-\frac{1}{2N} \sum_{i\neq j}J_{ij}M_{i}M_{j}\left\{\cos\left[\textit{\textbf{q}}\cdot \left(\textit{\textbf{R}}_{i}-\textit{\textbf{R}}_{j}\right)\right]\right.\nonumber\\
		& &\left.\times\sin\theta_{i}\sin\theta_{j}+\cos\theta_{i}\cos\theta_{j}\right\},
		\label{Heisenberg} 
\eeq
where $i\neq j$ is imposed to avoid the double counting of sites and $N$ denotes the number of unit cells. Transforming the exchange coupling parameter $J_{ij}$ in the real space to the reciprocal space
 \beq
J_{\textit{\textbf{q}}}=-\frac{1}{N}\sum_{i\neq j}J_{ij}e^{i\textit{\textbf{q}}\cdot\left(\textit{\textbf{R}}_{i}-\textit{\textbf{R}}_{j}\right)}.\label{exchange}
\eeq
and setting the small $\theta$, the magnon energy for one magnetic atom in the unit cell can be given as \cite{7Rosengaard, 10Kubler, 16Teguh, 23Sasioglu, Jakobson}
\beq
\hbar\omega_{\textit{\textbf{q}}}=\lim_{\theta\rightarrow 0} \frac{4\mu_{B}}{M}\frac{\Delta E(\textit{\textbf{q}},\theta)}{\sin^{2}\theta}.\label{magnonf}
\eeq 
Later, the plotted magnon energies close to $\textit{\textbf{q}}=0$ will be used to evaluate the spin stiffness constants, as will be discussed thoroughly in the next section.

The Curie temperature can be computed in two ways, using either the MFA or the RPA. In this case, the Curie temperature can be evaluated by the arithmetic average value of the magnon energies in MFA or by the harmonic average value of the magnon energies in RPA \cite{23Sasioglu}. Due to these different treatments, the MFA tends to give a larger value than the RPA if using the same number of discrete $\textit{\textbf{q}}$ point sampling. Although those two approaches can be used to estimate the Curie temperature, the RPA usually gives the Curie temperature more accurate than the MFA. The corresponding MFA and RPA can be formulated in the average value of the magnon energies as \cite{1Padja}
\beq
k_{B}T_{C}^{MFA}=\frac{M}{6\mu_{B}}\frac{1}{N}\sum_{\textit{\textbf{q}}}\hbar\omega_{\textit{\textbf{q}}}\label{MFA}
\eeq  
for the MFA and
\beq
\frac{1}{k_{B}T_{C}^{RPA}}=\frac{6\mu_{B}}{M}\frac{1}{N}\sum_{\textit{\textbf{q}}}\frac{1}{\hbar\omega_{\textit{\textbf{q}}}}\label{RPA}
\eeq  
for the RPA. Here, $M$ is the magnetic moment and $N$ is the number of the magnon energies. Note that there is a different view in taking $\textit{\textbf{q}}=0$, at which the summation becomes discontinue in the RPA, as seen in Eq. (\ref{RPA}). Most authors \cite{Sandratskii1, Sasioglu, Sandratskii2, Sandratskii3, Essenberg} excluded $\textit{\textbf{q}}=0$ in the calculations while Padja et al. \cite{1Padja} included $\textit{\textbf{q}}=0$ using the Green function.
    
\section{Results}    
First of all, the spin stiffness constant will be confirmed by observing its convergence based on two parameters, i.e., the cutoff radius and the number of orbitals. To do so, in the OpenMX code we specified these parameters by the symbol, for example, Fe5.0-$s3p3d2$, which means that we used a 5.0 cutoff radius in the atomic units (a.u.) for Fe atom with a basis set constructed by three $s$, three $p$, and two $d$ orbitals. To get reliable results, the choice of the cutoff radius and the number of orbitals requires the sufficient $k$ point. To handle this problem, we used a $50 \times 50 \times 50$ \emph{k} point grid for all the calculations involving the variation of the cutoff radius and the number of orbitals. We also set cutoff energy of 300 Ryd and used the local spin density approximation (LSDA) proposed by Ceperley and Alder \cite{Ceperley}. Although the generalized gradient approximation (GGA) can be used, however, in our calculation the use of GGA is very difficult for the small cone angle. 

To start the calculation, a conical spin spiral configuration was constructed by fixing a 10$^{\circ}$ cone angle by applying the constraint scheme method. To obtain the magnon energy, we computed the total energy difference from the self-consistent calculations and mapped it onto the effective Heisenberg Hamiltonian, as formulated in Eq. (\ref{magnonf}). We then varied the cutoff radius and increased the number of orbitals systematically. In this case, we stopped the number of orbitals, at which the overcompleteness appears, i.e., the total energy difference becomes unreliable. The spin stiffness constant $D$ will be then evaluated by a fourth-order fit $\hbar\omega_{\textit{\textbf{q}}}=Dq^{2}\left(1-\beta q^{2}\right)$.
\begin{figure*}[h!]
\vspace{2 mm}
\centering\includegraphics[scale=0.55, width =!, height =!]{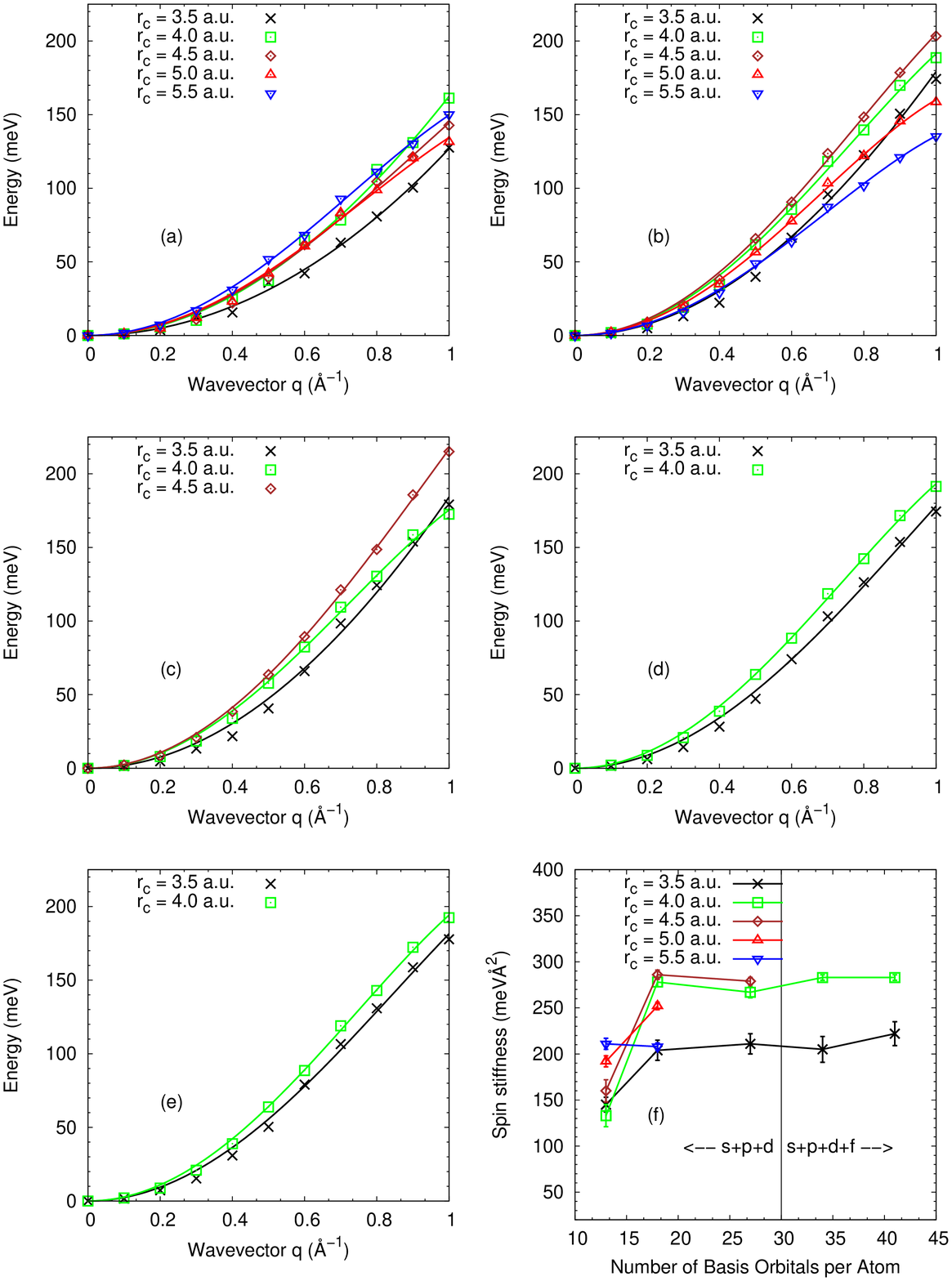}
\vspace{10mm}
\caption{\label{magnon_febcc}(Color online) Spectra of magnon energy near $\textit{\textbf{q}}=0$ in the (100) direction of bcc-Fe for some varied cutoff radii $r_{c}$ and the basis orbitals of (a) $s2p2d1$, (b) $s2p2d2$, (c) $s3p3d3$, (d) $s3p3d3f1$, (e) $s3p3d3f2$, respectively, where the fitting functions $\hbar\omega_{\textit{\textbf{q}}}=Dq^{2}(1-\beta q^{2})$ are shown by the solid lines. The dependence of spin stiffness constants on the number of orbitals is shown in (f). Some data have been reused from our last paper\cite{16Teguh}.} 
\end{figure*}

\begin{figure*}[h!]
\vspace{2mm}
\centering\includegraphics[scale=0.55, width =!, height =!]{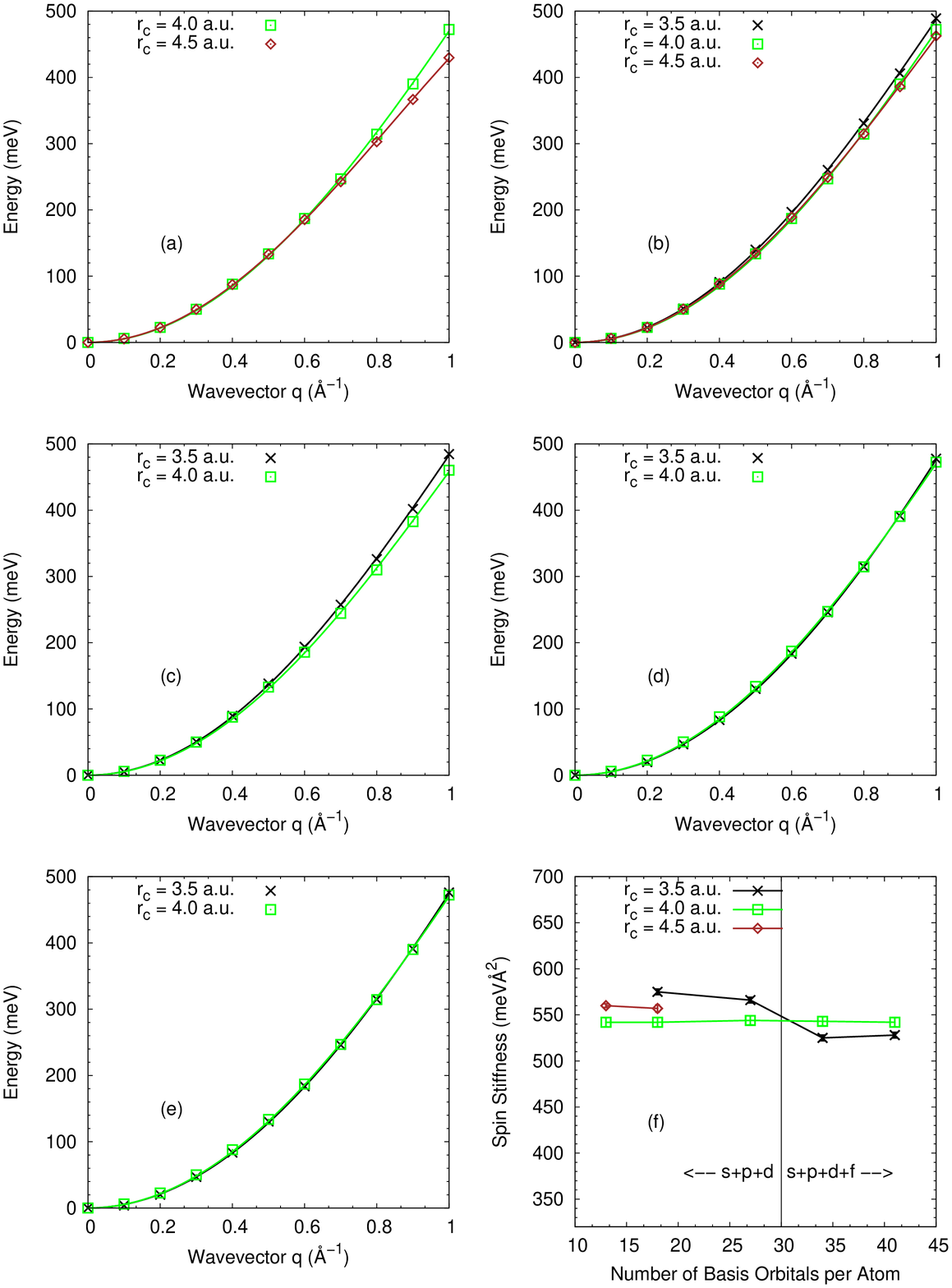}
\vspace{8mm}
\caption{\label{magnon_cofcc}(Color online) Spectra of magnon energy near $\textit{\textbf{q}}=0$ in the (001) direction of fcc-Co for some varied cutoff radii $r_{c}$ and the basis orbitals of (a) $s2p2d1$, (b) $s2p2d2$, (c) $s3p3d3$, (d) $s3p3d3f1$, (e) $s3p3d3f2$, respectively, where the fitting functions $\hbar\omega_{\textit{\textbf{q}}}=Dq^{2}(1-\beta q^{2})$ are shown by the solid lines. The dependence of spin stiffness constants on the number of orbitals is shown in (f).} 
\end{figure*}

\begin{figure*}[h!]
\vspace{2mm}
\centering\includegraphics[scale=0.55, width =!, height =!]{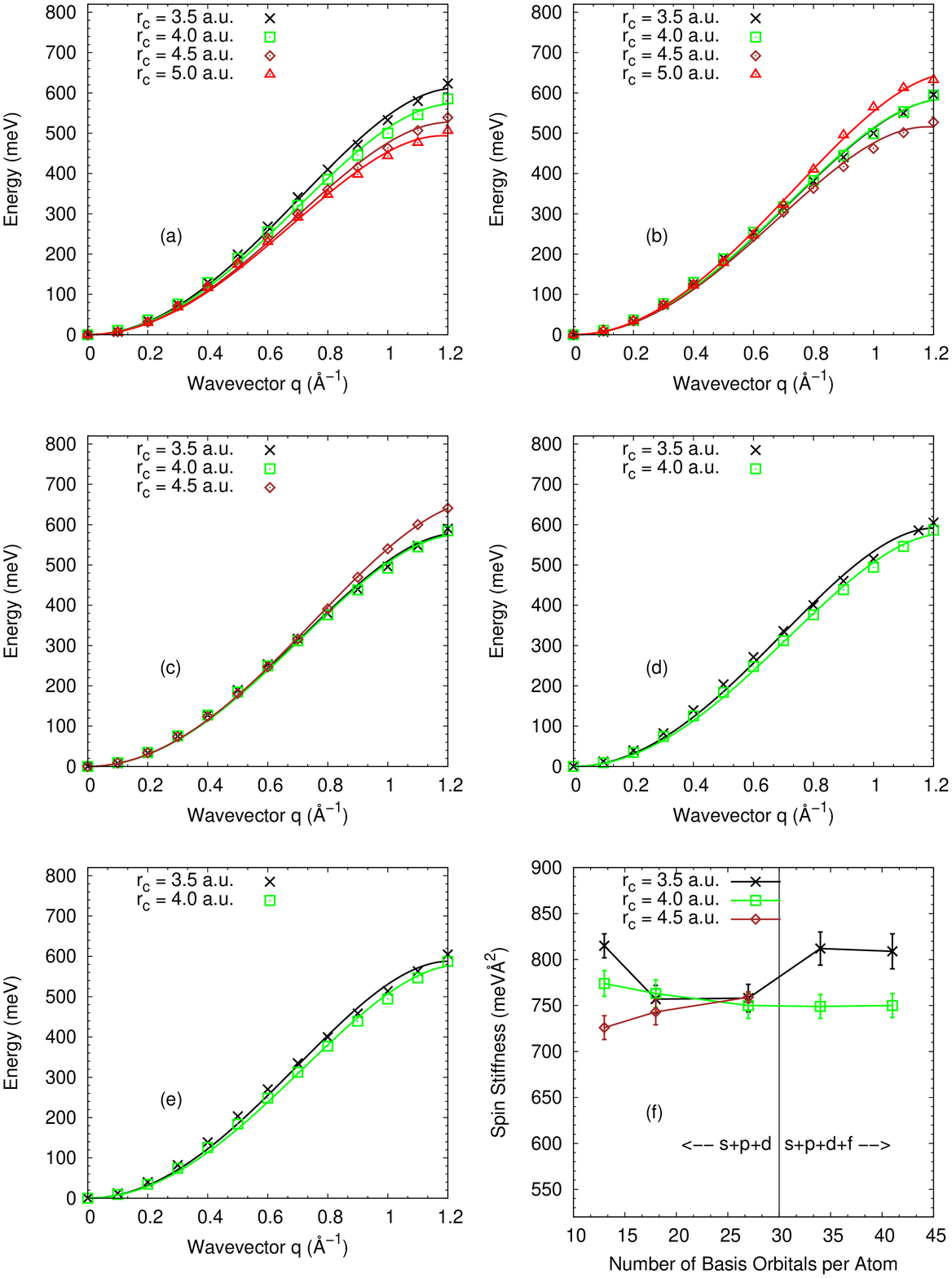}
\vspace{8mm}
\caption{\label{magnon_nifcc}(Color online) Spectra of magnon energy near $\textit{\textbf{q}}=0$ in the (001) direction of fcc-Ni for some varied cutoff radii $r_{c}$ and the basis orbitals of (a) $s2p2d1$, (b) $s2p2d2$, (c) $s3p3d3$, (d) $s3p3d3f1$, (e) $s3p3d3f2$, respectively, where the fitting functions $\hbar\omega_{\textit{\textbf{q}}}=Dq^{2}(1-\beta q^{2})$ are shown by the solid lines. The dependence of spin stiffness constants on the number of orbitals is shown in (f).} 
\end{figure*}

Figures \ref{magnon_febcc}-\ref{magnon_nifcc} show the magnon energies ((a)-(e)) and the convergences of spin stiffness constant (f) for bcc-Fe, fcc-Co, and fcc-Ni, respectively. For all figures, the number of orbitals can only be increased at the short cutoff radius, e.g., 3.5 a.u. or 4.0 a.u., while at the long cutoff radius, such as 5.0 a.u., the small number of orbitals can only be applied due to overcompleteness. Note that although the number of orbitals can be increased at the short cutoff radii, the small number of orbitals sometimes cannot be employed due to the insufficient basis set, as shown in Fig. \ref{magnon_cofcc}(a) for fcc-Co, in which the number of orbitals of $s2p2d1$ (11 orbitals) was not employed at the cutoff radius of 3.5 a.u. The spin stiffness constants for each cutoff radius and number of orbitals will then be plotted to see the convergence. We show that all the values of spin stiffness converge at the cutoff radius of 4.0 a.u. with the orbitals of $s3p3d3f2$ (41 orbitals), as shown in Figs. \ref{magnon_febcc}(f), \ref{magnon_cofcc}(f), and \ref{magnon_nifcc}(f). This suggests that the calculations of Curie temperature should be performed using these parameters.

The calculated spin stiffness constants for bcc-Fe, fcc-Co, and fcc-Ni using the cutoff radius of 4.0 a.u. with the orbitals of $s3p3d3f2$ are given in Table \ref{spin_stiffness}. We show that all the spin stiffness constants are very good agreement with the previous calculations. For bcc-Fe and fcc-Co, our results are very close to the experiment, but fcc-Ni has a deviation. This deviation has been regarded due to the role of Stoner excitations \cite{10Kubler}. Interestingly, Shallcross et al. obtained the accurate spin stiffness constant for fcc-Ni using the ferromagnetic magnetic force theorem (FM-MFT), but the disordered local moment magnetic force theorem (DLM-MFT) yielded a strong deviation for the spin stiffness constant of fcc-Ni \cite{9Shall}.       

\begin{figure*}[h!]
\vspace{2mm}
\centering\includegraphics[scale=0.55, width =!, height =!]{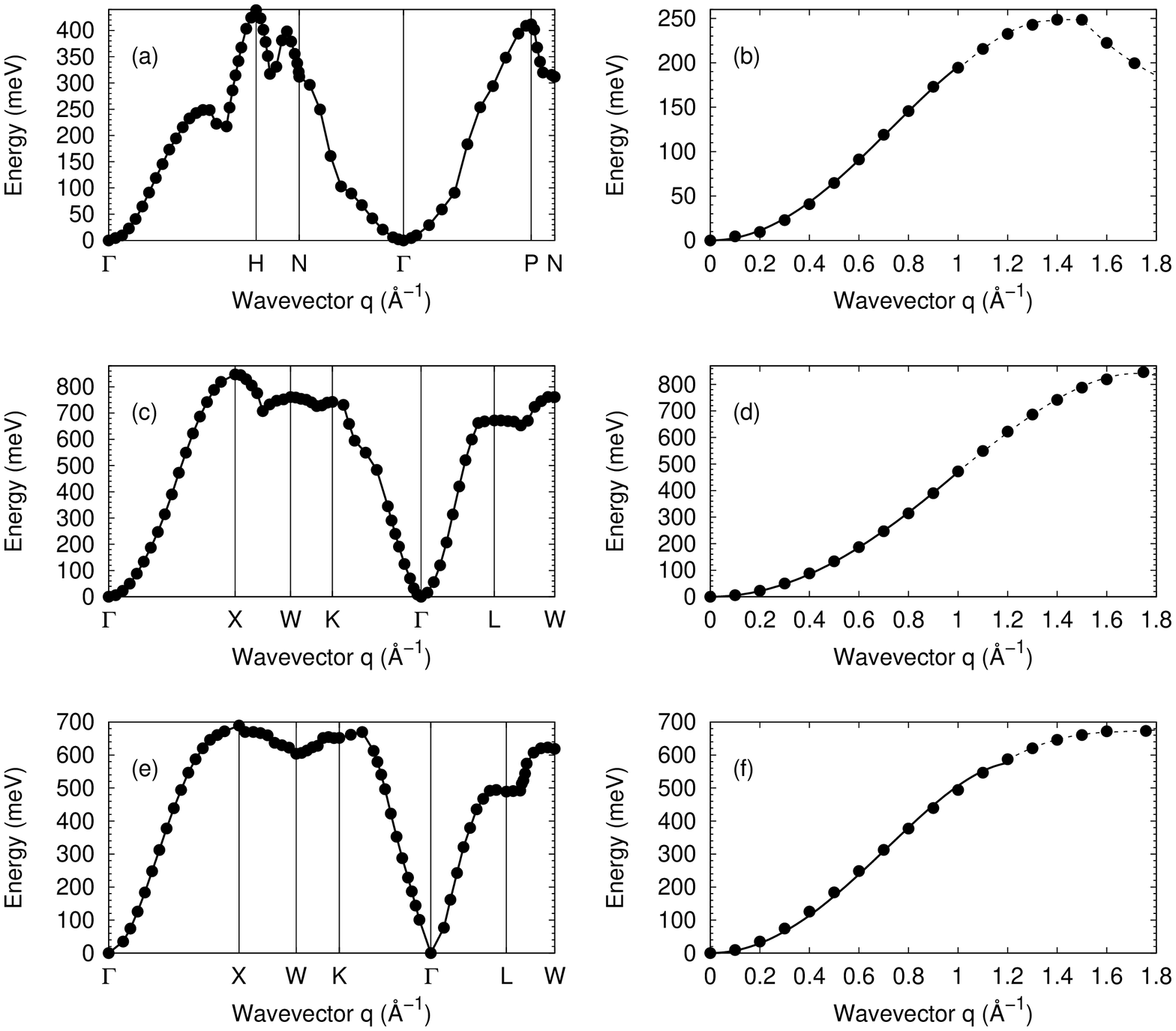}
\vspace{6mm}
\caption{\label{magnon_dispersion} Spectra of magnon energy using the cutoff radius of 4.0 a.u. and the orbitals of $s3p3d3f2$ on the high symmetry line in the Brillouin zone for bcc-Fe (a), fcc-Co (c), and fcc-Ni (e). The sets of the magnon energies for applying the Debye approximation are shown for bcc-Fe (b), fcc-Co (d), and fcc-Ni (f).} 
\end{figure*}

To support our LCPAO, we also provide the spectra of magnon energy on the high symmetry line in the Brillouin zone for bcc-Fe, fcc-Co, and fcc-Ni, as shown in Figs. \ref{magnon_dispersion}((a), (c), and (e)). The tendencies of our calculated magnon spectra are in good agreement with those using the LMTO method with the real space approach \cite{1Padja} and the frozen magnon approach \cite{8Halilov2}. Our results are only different on the peaks of the magnon energy in the special $k$ point. The parabolic curves, which are nearly isotropic, are observed in the long wavelengths near $\Gamma$ point for fcc-Co and fcc-Ni. For the short wavelengths in bcc-Fe, we also observe the so-called Kohn anomalies shown by two local minima in the interval of $\Gamma-\textrm{H}$ and $\textrm{H}-\textrm{N}$, in good agreement with Refs. \citen{1Padja, 8Halilov2}.   
 
We used two approaches for calculating the Curie temperatures. The first approach is to apply the MFA by taking the average of magnon energies in the Brillouin zone, as formulated in Eq. (\ref{MFA}). In general, our calculated results are in good agreement with the previous calculations, as shown in Table \ref{Curie_temperature}. Our estimation for bcc-Fe is also in good agreement with the experiments, whereas the deviation less than 15\% is addressed to fcc-Co and fcc-Ni. Note that, although our spin stiffness constant of fcc-Ni is overestimated from the experiment, as shown in Table \ref{spin_stiffness}, we obtain a Curie temperature closer to the experiment that those with the different methods in Refs. \citen{1Padja, 4Staunt, 5Uhl, 8Halilov2, 9Shall}.      
\begin{table}[ht]
\vspace{2 mm}
\caption{Evaluated spin stiffness constants $D$ (meV{\AA}$^{2}$) using both first-principles calculations and experimental results. Here, the cutoff radius of 4.0 a.u. and the orbitals of $s3p3d3f2$ are used.}   
\centering 
\begin{tabular}{c c c c c c c} 
\hline 
Metal&$D_{cal}^{1}$&$D_{cal}^{2}$&$D_{cal}^{3}$&$D_{cal}^{4}$&$D_{cal}^{5}$&$D_{exp}$\\ 
\hline 
bcc-Fe&283&250&247&322, 313&355&314$^{a}$, 230$^{b}$, 280$^{c}$, 307$^{d}$\\
fcc-Co&542&663&502&480, 520&535&510$^{e}$, 580$^{c}$\\
fcc-Ni&750&756&739&541, 1796& 715&422$^{c}$, 550$^{f}$, 555$^{g}$\\
\hline 
\end{tabular}
\begin{tablenotes}
  \item[1] \footnotesize $^{1}${Present calculation.} 
  \item[2] $^{2}$Calculation from Ref. \citen{1Padja}. 
	\item[3] $^{3}$Calculation from Ref. \citen{7Rosengaard}. 
  \item[4] $^{4}$Calculation from Ref. \citen{9Shall}. 
	\item[5] $^{5}$Calculation from Ref. \citen{10Kubler}.
  \item[a] $^{a}$Taken from Ref. \citen{Stringfellow}.
	\item[b] $^{b}$Taken from Ref. \citen{Lynn}.
	\item[c] $^{c}$Taken from Ref. \citen{Pauthenet}. 
  \item[d] $^{d}$Taken from Ref. \citen{Loong}. 
	\item[e] $^{e}$Taken from Ref. \citen{Shirane}.
  \item[f] $^{f}$Taken from Ref. \citen{Mitchell}.
	\item[g] $^{g}$Taken from Ref. \citen{Mook}.
  \end{tablenotes}
\label{spin_stiffness} 
\vspace{-6 mm}
\end{table}

The second approach to calculate the Curie temperature is addressed to the RPA. Since the number of $\textit{\textbf{q}}$ points in the Brillouin zone is required much denser than that of the MFA, we consider the so-called Debye approximation. The main notion of this approximation is to replace the discrete calculation in Eq. (\ref{RPA}) with the integration over a sphere with the defined radius as if it is performed within the Brillouin zone. This means that the sphere with radius $q_{D}=(6\pi^{2}/\Omega)^{1/3}$, where $\Omega$ is the volume of the unit cell, can represent the Brillouin zone with the same volume. This approximation works well for some Heusler alloys for calculating the Curie temperature \cite {Enkovaara,Lezaic}.  
\begin{table*}[h]
\vspace{2 mm}
\caption{Evaluated Curie temperatures $T_{C}$ (Kelvin) using both first-principles calculations and experiments. Here, the cutoff radius of 4.0 a.u. and the orbitals of $s3p3d3f2$ are used.}   
\centering 
\begin{tabular}{c c c c c c c c c c c} 
\hline 
Metal&$T_{C}^{1}$&$T_{C}^{2}$&$T_{C}^{3}$&$T_{C}^{4}$&$T_{C}^{5}$&$T_{C}^{6}$&$T_{C}^{7}$&$T_{C}^{8}$&$T_{C}^{9}$&$T_{C}^{exp}$\\ 
\hline 
bcc-Fe&1013, 742&1414, 950&1270&1015&1095&1460, 1060&1037&550, 1190&1316&1043-1045\\
fcc-Co&1534, 1422&1645, 1311&1520&&1012&1770, 1080&1250&1120, 1350&1558&1388-1398\\
fcc-Ni&555, 547&397, 350&&450&412&660, 510&430&320, 820&642&624-631\\
\hline 
\end{tabular}
\begin{tablenotes}
  \item[1] \footnotesize $^{1}${Present calculation. The left and right sides were calculated using the MFA and the RPA, respectively} 
  \item[2] $^{2}$Calculation from Ref. \citen{1Padja}. 
	\item[3] $^{3}$Calculation from Ref. \citen{Sabiryanov}. 
	\item[4] $^{4}$Calculation from Ref. \citen{4Staunt}. 
	\item[5] $^{5}$Calculation from Ref. \citen{5Uhl}. 
	\item[6] $^{6}$Calculation from Ref. \citen{7Rosengaard}.   
  \item[7] $^{7}$Calculation from Ref. \citen{8Halilov2}. 
	\item[8] $^{8}$Calculation from Ref. \citen{9Shall}. 
	\item[9] $^{9}$Calculation from Ref \citen{10Kubler}. 
  \end{tablenotes}
\label{Curie_temperature} 
\vspace{-6 mm}
\end{table*}

To apply the Debye approximation, we initially modify Eq. (\ref{RPA}) in the integral formulation \cite {Enkovaara,Lezaic}
\beq
\frac{1}{k_{B}T_{C}^{RPA}}=\frac{6\mu_{B}}{M}\frac{\Omega}{2\pi^{3}}\int d^{3}{\textit{\textbf{q}}}\frac{1}{\hbar\omega_{\textit{\textbf{q}}}}\label{Debye},
\eeq  
where the magnon energy $\hbar\omega_{\textit{\textbf{q}}}$ is replaced by the appropriate fitting function, as carried out for calculating the spin stiffness constant. Here, the radii $q_{D}$ for bcc-Fe, fcc-Co, and fcc-Ni are found to be 1.712 {\AA}$^{-1}$, 1.748 {\AA}$^{-1}$, and 1.758 {\AA}$^{-1}$, respectively. The schematic calculations using Eq. (\ref{Debye}) can be seen in Figs. \ref{magnon_dispersion}((b), (d), and (f)). The integral calculation is divided into two steps. First, one makes the integration from $q=0$ to $q$, at which the spin stiffness constant can be obtained, for the explanation see the solid lines in Figs. \ref{magnon_dispersion}((b), (d), and (f)). At last, one should continue the integration by finding the appropriate fitting function, see the dashed lines in Figs. \ref{magnon_dispersion}((b), (d), and (f)). From Table \ref{Curie_temperature}, we see the better results of the calculated Curie temperature, except for the bcc-Fe, for which the Curie temperature is underestimated so largely. 

We would like to give some comments on why the Curie temperature is underestimated for bcc-Fe. By seeing the dashed lines in Figs. \ref{magnon_dispersion}((b), (d), and (f)), we suppose that the underestimate or the overestimate depends on the tendency of the fitting function. The calculated Curie temperature for fcc-Co gets the error of less than 5 \%, which is the best result, followed then by fcc-Ni and bcc-Fe. In this case, we see that the dashed line for fcc-Co follows the flow of the solid line, which means that there is almost no deviation, see Fig. \ref{magnon_dispersion}(d). For fcc-Ni, we see a small deviation, especially at $q \geq 1.6$ {\AA}$^{-1}$ that forms a straight line, see Fig. 4(f). The large deviation is observed for bcc-Fe at the range of $q \geq 1.5$ {\AA}$^{-1}$, which is drastically decreased, see Fig. 4(b). This means that at that range the magnon energies enter the Kohn anomalies region. Therefore, if we want to improve the result, we should integrate only up to $q=1.5$ {\AA}$^{-1}$, which gives 994 K, closer to the experiment than the previous one.

\section{Conclusions}
We have presented a schematic procedure to calculate the spin stiffness constants, as well as the Curie temperatures by using the GBT with an LCPAO as the basis set. The convergences of spin stiffness constant of bcc-Fe, fcc-Co, and fcc-Ni have been determined by two parameters, the cutoff radius and the number of orbitals. These convergences can only be achieved by choosing the cutoff radius of 4.0 a.u. with the orbitals of $s3p3d3f2$, which gives 41 orbitals. We also show the strong dependence between the cutoff radius and the number of orbitals. This dependence determines how many orbitals can be set in the cutoff radius. For the short cutoff radius, the small number orbitals, in general, is not sufficient to represent the wavefunctions. On the contrary, if a large number of orbitals is set to the long cutoff radius, the overcompleteness appears. All the calculated spin stiffness constants are in good agreement with both other methods and experiments, except for fcc-Ni that yields a large deviation. 

The spectra of the magnon energy and the Curie temperatures are derived by setting the cutoff radius of 4.0 a.u. with the orbitals of $s3p3d3f2$, at which the calculated spin stiffness constants converge. All the tendencies of the magnon spectra for all systems are in good agreement with the other methods. The Curie temperatures are estimated by means of two ways, the MFA and the RPA. The average of magnon energies in the Brillouin zone has been used to calculate the Curie temperature using the MFA, by which our estimations are in good agreement with the experiments. On the other side, we apply the so-called Debye approximation to estimate the Curie temperatures with the RPA. Although using the very small number of $\textit{\textbf{q}}$ points in this approximation, we prove that the calculated Curie temperatures can also be in good agreement with the experiments as long as the appropriate $q_{D}$ is chosen.              
                     
\section*{Acknowledgments}
T. B. P. wishes to thank Dr. M. Le$\breve{\textrm{z}}$a{\'{i}}c for the available discussion on the Debye approximation. The computations were carried out using ISSP supercomputers located at the University of Tokyo, while the spectra of magnon energy and calculated Curie temperatures were performed at the Universitas Negeri Jakarta.

\end{document}